\documentclass[aps,prb,twocolumn,showpacs,superscriptaddress,
groupedaddress]{revtex4-1}  

\usepackage[english]{babel}
\makeatletter
\adddialect\l@en\l@english
\makeatother

\usepackage{graphicx}  
\usepackage{amssymb}   
\usepackage{natbib}    
\usepackage{color}     
\usepackage{booktabs} 
\usepackage{bm}
\usepackage[normalem]{ulem}

\begin{document}

\newcommand{\ve}[1]{\mathbf{#1}}
\newcommand{\ket}[1]{\left| #1 \right\rangle}
\newcommand{\bra}[1]{\left\langle #1 \right|}
\newcommand{\braket}[3]{\left\langle #1 \left| #2 \right| #3 \right\rangle}
\newcommand{\F}{$\mathcal{\bf F}$:}

\title{Electron-correlation driven capture and release in double 
quantum dots}
\author{Federico M. Pont}
\email{pont@famaf.unc.edu.ar}
\affiliation{Facultad de Matem\'atica, Astronom\'{i}a y F\'{i}sica, Universidad Nacional de C\'ordoba, and IFEG-CONICET, Ciudad Universitaria, X5000HUA, C\'{o}rdoba, Argentina
and Theoretische Chemie, Physikalisch-Chemisches Institut, Im Neuenheimer Feld 229, 69120 Heidelberg, Germany}
\author{Annika Bande}
\email{annika.bande@helmholtz-berlin.de}
\affiliation{Institute of Methods for Material Development and Joint Ultrafast Dynamics Lab
in Solutions and at Interfaces (JULiq),
Helmholtz-Zentrum Berlin f\"ur Materialien und Energie, Albert-Einstein-Str. 15,
12489 Berlin, Germany}
\affiliation{Theoretische Chemie, Physikalisch-Chemisches Institut, Im Neuenheimer Feld 229, 69120 Heidelberg, Germany}
\author{Lorenz S. Cederbaum}
\email{lorenz.cederbaum@pci.uni-heidelberg.de}
\affiliation{Theoretische Chemie, Physikalisch-Chemisches Institut, Im Neuenheimer Feld 229, 69120 Heidelberg, Germany}

\vskip 0.25cm
\date{\today}

\begin{abstract}
We recently predicted that the interatomic Coulombic 
electron capture (ICEC) process, a long-range electron 
correlation driven capture process, is achievable in
gated double quantum dots (DQDs). In ICEC an incoming 
electron is captured by one QD and the excess 
energy is used to remove an electron from the neighboring QD. In this work we 
present systematic full three-dimensional electron dynamics calculations
in quasi-one dimensional model potentials that allow for a detailed understanding of the 
connection between the DQD geometry and the reaction probability for the 
ICEC process. We derive an effective one-dimensional approach and show that its results compare very well with
those obtained using the full three-dimensional calculations. This approach substantially reduces the
computation times. The investigation of the electronic structure for various DQD geometries for which the ICEC process can
take place clarify the origin of its remarkably high probability in the presence of two-electron
resonances.

\end{abstract}

\maketitle

\section{Introduction}

The technical ability of producing nanosized materials lead among other achievements to the 
discovery - and nowadays the technological application 
~\cite{ameenah_alahmadi_quantum_2012} - of semiconductor (SC) QDs. In these 
structures some typical features of SC bulk material are 
prevailed~\cite{fujisawa_spontaneous_1998,
shabaev_multiexciton_2006, muller_electrical_2012, benyoucef_single-photon_2012} 
and married to typical atomic properties~\cite{van_der_wiel_electron_2002, 
salfi_electronic_2010,
laird_coherent_2010, roddaro_manipulation_2011, nadj-perge_spectroscopy_2012} 
emerging from the energy level quantization~\cite{reed_observation_1988} in the 
QDs, motivating their name: artificial atoms.~\cite{kastner_artificial_1993} DQDs 
can either be coupled (artificial molecules~\cite{van_der_wiel_electron_2002}) 
or uncoupled. The latter arrangement we consider here for the investigation of 
an energy transfer process between QDs.

The electron confinement achieved through
different QD geometries (disc 
shaped, spherical, wires, double layered, etc.) presents an interesting 
variety of electronic properties that are, however, similar for various kinds 
of QDs. Epitaxially-grown self-assembled QDs are most commonly disc or 
pyramidally shaped InGaAs islands onto a GaAs substrate fed through a wetting layer by free 
electrons from the substrate.~\cite{goldstein_growth_1985, henini_handbook_2011}
Vertical stacking of layers allows to obtain a nanostructure
of vertically arranged DQDs.~\cite{goldstein_growth_1985, henini_handbook_2011}

In electrostatically defined QDs, a two-dimensional 
electron gas is created between two semiconductors with different gaps. 
The gas can carry free electrons which can be further 
confined using charged metallic
gates to define the regions of one, 
two or more QDs.~\cite{van_der_wiel_electron_2002} 
 In the last years the advances in nanowire fabrication allowed the 
construction of QDs inside long nanowires using interlaced layer of
different semiconductors.~\cite{salfi_electronic_2010} 
Colloidal nanocrystals can nowadays be constructed small enough
to observe quantization of the electronic levels. They have
attracted a lot of attention in the past few years as materials in modern third 
generation solar cells.~\cite{gur_air-stable_2005,nozik_semiconductor_2010} 
In all theses QD structures the manipulation of the electronic levels 
of the QDs is straightforward. 
Particularly, manipulation of levels with 
different spin quantum numbers
by 
magnetic or electric fields is possible. This allows the study and 
characterization of transitions between 
them,~\cite{roddaro_manipulation_2011,salfi_electronic_2010, 
fujita_nondestructive_2013, studenikin_quantum_2012, muller_electrical_2012, 
porte_ultrafast_2009, nadj-perge_spectroscopy_2012} which are an appealing and desirable 
property in the field of quantum information. 

Many experimental techniques are employed in current research to measure the 
properties of QDs. The electrical current through QDs can be obtained by 
transport spectroscopy. Transport on electrostatically 
defined QDs,~\cite{van_der_wiel_electron_2002} nanowire based 
QD structures,~\cite{salfi_electronic_2010,roddaro_manipulation_2011} 
and nanotube defined QDs~\cite{leturcq_franckcondon_2009} 
is widely used to determine the level structure inside the QDs. Another
important field of research in 
various nanostructures is carrier relaxation dynamics within excitons after 
an optical excitation. Pump-probe schemes with time resolution in the order of ten of picoseconds
can resolve processes such as electron-phonon
interactions,~\cite{prasankumar_ultrafast_2009, porte_ultrafast_2009,
zibik_long_2009} multiple exciton generation,~\cite{nozik_semiconductor_2010}
Auger relaxation~\cite{narvaez_carrier_2006} also far-IR relaxation and relaxation into defects,
impurities especially at surfaces. The characteristics can be measured by photoluminescence
spectroscopy~\cite{muller_electrical_2012, benyoucef_single-photon_2012,
shirasaki_emergence_2013} and complementary photocurrent measurements can give 
information on the non-radiative decay time and energy of the excitons or
intra-conduction band excited states.~\cite{muller_electrical_2012} In 
the specific case of DQDs, the transitions and tunneling dynamics of electrons 
of vertically coupled QDs were studied~\cite{muller_electrical_2012}
and interdot phonon-relaxation processes were detected between the QDs.
P to S orbital electron relaxation via electron correlation has also been 
demonstrated in uncoupled $n$-doped DQDs~\cite{bande_dynamics_2011, bande_electron_2013, cherkes_electron_2011} and 
after electric pulse excitation.~\cite{bande_electron_2013-1} In this case the relaxation in 
one QD occurs via energy transfer and emission of an electron in a neighboring 
QD in a process called intermolecular Coulombic decay 
(ICD).~\cite{bande_dynamics_2011, cherkes_electron_2011, cederbaum_giant_1997, 
sisourat_interatomic_2010, sisourat_ultralong-range_2010, 
jahnke_ultrafast_2010}

\begin{figure}
\includegraphics[width=0.45\textwidth]{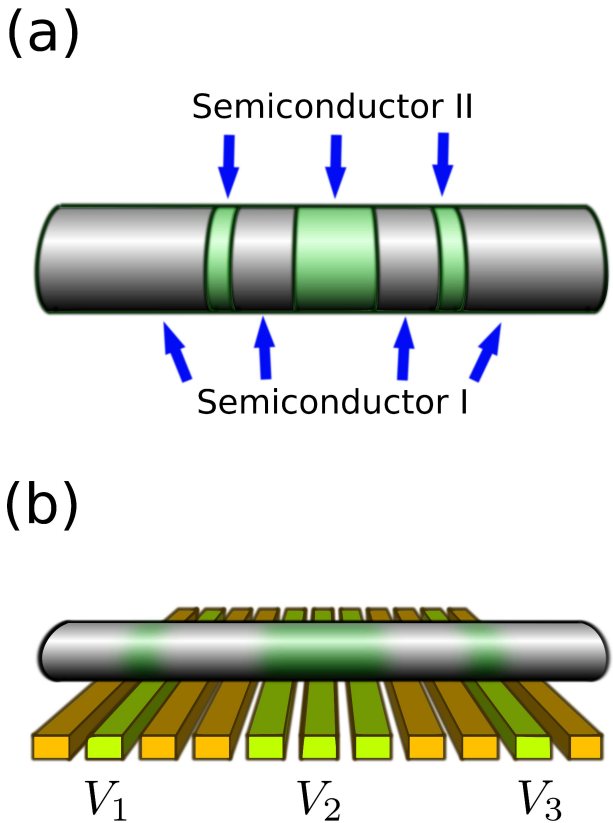}
\caption{\label{types_of_nanowires}(Color online) Schematic view of two 
experimental setups to 
achieve the electron confinement inside a nanowire. In panel (a) a 3D 
confinement is obtained using a layered semiconductor structure, in (b) the 
nanowire is built of a single semiconductor material and
the barriers are obtained by electrostatic depletion (areas indicated with
green shading). The depletion is achieved by setting different electrostatic 
potential energies in the metallic gates below the wire.}
\end{figure}

In the present work we focus on the less intensively studied
capture dynamics of free electrons into n-doped DQDs mediated solely 
by long-range electron correlation.~\cite{pont_controlled_2013} 
In general the most important 
electron capture mechanism is via emission of 
longitudinal optical phonons, that has been studied before in 
single~\cite{glanemann_transport_2005, jiang_inelastic_2012} and double
QDs.~\cite{glanemann_transport_2005} It has been analyzed 
theoretically in single QDs along with electron 
collisions and emission.~\cite{glanemann_transport_2005, 
kvaal_multiconfigurational_2011} In our 
previous work~\cite{pont_controlled_2013} we showed for the first 
time that electron capture can as well be mediated 
efficiently by long-range electron correlation in the interatomic Coulombic 
electron capture (ICEC) in DQDs. The process was named
after the one originally predicted to be operative
in atoms and molecules.~\cite{gokhberg_environment_2009, 
gokhberg_interatomic_2010} 
In atoms the electron capture by one atom occurs while another
electron is emitted from an atom into its environment.
In DQDs the electron capture by one QD leads to an emission of electrons from
neighboring QDs with controlled energy properties that can be tuned by changing 
the geometric DQD
parameters.~\cite{pont_controlled_2013} 
We postulated ICEC for n-doped DQDs embedded in nanowires (Fig.~\ref{types_of_nanowires}) 
using an effective mass approximation (EMA)~\cite{bastard_wave_1991} based model potential in which we 
performed numerically exact electron dynamics calculations. The 
relaxation dynamics of an excitonic electron in undoped materials can be described within 
the same model provided that the hole relaxation to the band edge has been faster 
than that of the electron.~\cite{narvaez_carrier_2006}

We showed already that the probability for ICEC is non-negligible 
~\cite{pont_controlled_2013} and can be greatly enhanced in the 
presence of two-electron resonance states that are capable of 
undergoing fast ICD-related energy transfer. Here, we systematically 
add other DQD configurations to those studied before and analyze 
how and for which energies in the different configurations ICEC in 
the general and the resonance case becomes most effective. 

The paper is organized as follows: First we present some general 
considerations on the ICEC process~(\ref{general}), introduce our model and the 
DQD electronic structure~(\ref{model}) followed by the electron dynamics 
methods used~(\ref{methods}) and the results~(\ref{results}).
Since numerically exact computations in the full six-dimensional
Hilbert space are very time consuming, we additionally include an effective
two-dimensional description of the nanowires and compare to the full dimensional
results (\ref{1dicec}). The discussion of the results using realistic semiconductor
parameters are given in~(\ref{discussion}) followed by the conclusions~(\ref{conclusion}).

\section[general]{Conditions for ICEC in DQDs}\label{general}

\begin{figure}
\includegraphics[width=0.49\textwidth]{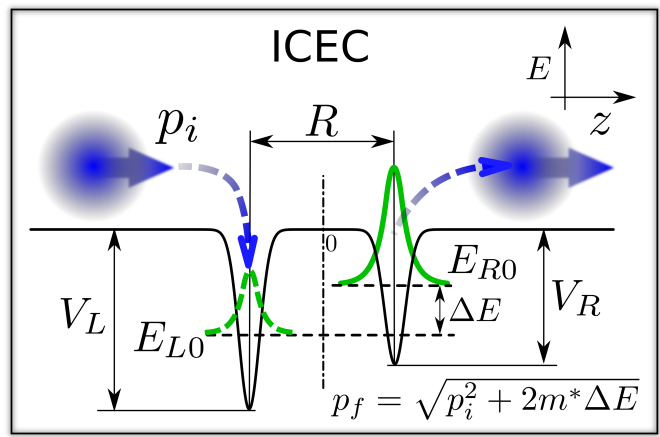}
\caption{\label{scheme}(Color online) Schematic view of the interatomic Coulombic
electron capture for a double quantum dot. The effective mass approximation is used to describe 
the quantum dots as two potential wells. The capture of the incoming electron by the left dot
(dashed green state) is mediated by its correlation with the electron initially 
bound to the right dot (full green state). While the electron is captured in 
the left dot, the electron on the right is excited into the continuum and 
becomes an outgoing electron.}
\end{figure}

In this work we consider a system of two fully correlated electrons and two 
QDs which we call the left and right QD and which are described 
by two different model potentials (see Fig.~\ref{scheme}). 
For the time being consider a left potential well that supports only a 
single one-electron level $L_0$ with energy $E_{L_0}$ and a right one with 
one single-electron level $R_0$ with energy $E_{R_0}$ such that $E_{L_0}\neq E_{R_0}$.
The tunneling and hybridization between $L_0$ and $R_0$ 
in the DQD is vanishingly small due to the long interdot distance 
$R$ of the considered system. The ICEC process occurs as depicted in 
Fig.~\ref{scheme} where
an electron is initially bound to the right QD and 
another electron with momentum $p_i$ is
coming in from the left side of the DQD. The 
incoming electron can then be captured
into the $L_0$ ground state of the left QD while the electron on the right is
emitted from the $R_0$ ground state of the right QD. 
Energy conservation dictates that the total energy 
of the system $E_T$ 
\begin{eqnarray}
\textrm{(in)}\quad E_T &=&  \varepsilon_{i} + E_{R_0} 
\label{total_energy_in}\\
\textrm{(out)}\quad E_T &=& \varepsilon_{f} + E_{L_0} \label{total_energy_out},
\end{eqnarray}

\noindent is conserved~\cite{gokhberg_interatomic_2010} and the kinetic energy acquired by the outgoing electron can be expressed as
\begin{equation}\label{energy_cons}
\varepsilon_{f}-\varepsilon_{i} = \Delta E
\end{equation}

\noindent with the corresponding momentum 
\begin{equation}\label{final_momentum}
p_f=\sqrt{p^2_i+2m^*\Delta E} 
\end{equation}

\noindent where $\varepsilon_{{i,f}}=p^2_{i,f}/2 m^*$, $\Delta 
E=E_{R_0}-E_{L_0}$ and $m^*$ is the electron effective
mass in atomic units. As one can notice from Eq.~(\ref{final_momentum}) the 
emitted electron can have a higher or a lower momentum than the initial 
electron, depending on the relation between the bound-state energies $E_{R_0}$ 
and $E_{L_0}$. However, for negative values of $\Delta 
E$ the ICEC channel is closed if the incoming electron energy is lower than 
$|\Delta E|$ (see Eq.~(\ref{final_momentum})). Note also that since 
$\Delta E$ is the energy acquired by the outgoing electron, then $-\Delta E$ is 
conversely
the energy gain/loss suffered by the DQD.

\section[model]{Model}\label{model}

The motion of two electrons inside a nanostructured semiconductor can be 
accurately described using a few-electron effective 
mass model potential~\cite{bastard_wave_1991} in which 
electron dynamics calculations are feasible. This approach offers then 
straightforward observability of how electron correlation
can lead to ICEC in general two-site systems where electron correlation between 
moieties plays a fundamental role as well as in the specific case of a QD. We 
adopt here the model for the DQD used previously to study the dynamics of 
ICEC~\cite{pont_controlled_2013,bande_interatomic_2015}
and 
ICD.~\cite{bande_dynamics_2011,bande_electron_2013,
bande_electron_2013-1} The dots are 
represented by two Gaussian wells aligned in $z$ direction. In $x$ and $y$ 
direction we assume a strong harmonic confinement which could be attributed 
either to depleting
gates~\cite{fujisawa_spontaneous_1998} or to the actual structure of the
semiconductor.~\cite{salfi_electronic_2010}
Besides the full three-dimensional calculations we also considered a 
simpler one-dimensional model that
uses an effective electron-electron interaction to take the wire shape of the 
system in $x$ and $y$ direction implicitly into account.
In this one-dimensional effective model electron dynamics 
calculations are much more efficient because only the $z$ coordinates of the 
electrons are evolved in time.

\subsection[hamiltonian]{Hamiltonian}\label{hamiltonian}
The two-electron effective mass Hamiltonian for the system is
\begin{equation}\label{ham}
H(\ve{r}_1, \ve{r}_2) = h(\ve{r}_1) + h(\ve{r}_2) +
\frac{1}{\varepsilon_r\left|\ve{r}_1-\ve{r}_2\right|}
\end{equation}

\noindent where $\varepsilon_r$ 
is the relative dielectric permittivity and
\begin{equation}\label{1eham}
h(\ve{r}_i) = -\frac{1}{2 m^*}\nabla_i^2 + V_{c}(x_i,y_i) +
V_{l}(z_i)
\end{equation}

\noindent is a one-electron Hamiltonian in which
\begin{eqnarray}
V_{c}(x_i,y_i) &=& \frac{1}{2}m^*\omega^2(x_i + y_i)^2 \label{vc} \\
V_{l}(z_i) &=& - V_L e^{-b_L (z_i + R/2)^2} - V_R e^{-b_R (z_i - R/2)^2}
\end{eqnarray}

\noindent are the transversal confinement and longitudinal open potentials,
respectively. $m^*$ is the effective mass, $R$ is the distance between the QDs 
and $b_{L,R}$ are the sizes of
the left and right QD while $V_{L,R}$ their depths. Performing the 
scaling $\ve{r}_i \rightarrow \frac{\varepsilon_r}{m^*}\ve{r}_i$ of the 
electronic coordinates one can obtain the scaling relationships of the 
Hamiltonian parameters shown in Tab.~\ref{tab:H_par}. Clearly, we can use the
effective mass and the relative permittivity equal to one and rescale the 
parameters afterwards to obtain the energies and distances for a specific semiconductor.

\begin{table}
\centering 
\begin{tabular}{ c | c } 
\hline 
Parameter & Scaled value \\
\hline
 & \\
$H$ (or $E$) & $\frac{m^*}{\varepsilon_r^2}$ $H$ \\
$m^*$ & $1$ \\
$\varepsilon_r$ & $1$ \\
$\omega$ & $\omega$ \\
$R$ & $\frac{\varepsilon_r}{m^*}\, R$\\
$(b_L,b_R)$ & $\frac{\varepsilon_r}{m^*}\, (b_L,b_R)$\\
$(V_L,V_R)$ & $\frac{\varepsilon_r^2}{m^*}\, (V_L,V_R)$\\
 & \\
\hline
\end{tabular}
\caption{Scaling of the Hamiltonian and parameters under the 
transformation $\ve{r}_i \rightarrow \frac{\varepsilon_r}{m^*}\ve{r}_i$.} %
\label{tab:H_par} 
\end{table}

Due to the comparably strong confinement ($\omega=1.0$ a.u. $> V_{L,R}$) the 
excited states relevant to this
study are only in $z$ direction. We will correspondingly have a level structure
$L_n(R_n)$, $n=0,1, \ldots$ in the left (right) QD with 
energies
$E_{L_n}(E_{R_n})$. The orbital symmetry is simply that of a symmetric well: $L_0$
corresponds to an S-symmetry around the left dot, $L_1$ to a P-symmetry and so on.

\subsection[1dmodel]{Effective one-dimensional approach}\label{1dmodel}
As mentioned in Sec.~\ref{hamiltonian} the system under consideration 
has a strong lateral confinement. It
is then possible to construct an
effective one-dimensional Hamiltonian~\cite{bednarek_effective_2003} using 
the wave function separation ansatz
\begin{equation}\label{ansatz}
 \Psi(\ve{r}_1,\ve{r}_2) = \psi(z_1,z_2)\phi_0(x_1,y_1)\phi_0(x_2,y_2),
\end{equation}

\noindent where $\phi_0$ are two-dimensional single-electron ground 
state functions and $\psi(z_1,z_2)$ is the longitudinal 
effective wave function. Since essentially the same 
results are obtained for singlet and triplet states, we chose 
triplet symmetry throughout our study. $\Psi$ has the proper symmetry under 
exchange of 
electrons given by the longitudinal wave 
function 
$\hat{\Pi}_{1\,\leftrightarrow\,2}\psi(z_1,z_2)=-\psi(z_2,z_1)$. The 
one-dimensional 
Hamiltonian can be deduced from the analysis of the expectation value of the 
full Hamiltonian with the product wave function of Eq. 
(\ref{ansatz})
\begin{eqnarray}\label{ansatztoham}
 \braket{\Psi}{H}{\Psi} &=&
2\omega -\sum_{i=1,2} \braket{\psi}{\frac{1}{2
m^{*}_i}\frac{\partial^2}{\partial z_i^2} +
V_{long}(z_i)}{\psi}+\nonumber\\
& 
&\frac{1}{\varepsilon_r}\braket{\Psi}{\frac{1}{\left|\ve{r}_1-\ve{r}
_2\right|}}{
\Psi } . 
\end{eqnarray}

The last term can be explicitly written in the form 
\begin{eqnarray}\label{intcoul}
&\braket{\Psi}{\frac{1}{|\ve{r}_{1}-\ve{r}_2|}}{\Psi}=&
\int \int
\frac{|\Psi(\ve{r}_1,\ve{r}_2)|^2}{| \ve{r}_1-\ve{r}_2 |} d \ve{r}^3_1 d 
\ve{r}^3_2 \nonumber \\
= \int \int &|\psi(z_1,z_2)|^2& V_{eff}(z_{12})  dz_1 dz_2,
\end{eqnarray}
\noindent with the squared longitudinal wave function and the 
effective $z$-potential
\begin{equation}
V_{eff}(z_{12})=
\sqrt{\frac{\pi}{2}}\frac{1}{l}e^{\zeta^2}\left(1-erf(\zeta)\right),
\end{equation}
\noindent which depends on $z_{12}=|z_1-z_2|$, the variable remaining after 
integrating over the $x$ and $y$ coordinates.

The size 
of the two-dimensional ground state wave function is given by 
$l=\sqrt{\braket{\phi_0}{x^2}{\phi_0}}=\sqrt{1/m^{*}\omega}$ and 
$\zeta= z_{12}/\sqrt{2}\, l$  is the distance $z_{12}$ between the electrons 
in terms of the confinement size $l$. The asymptotic behavior of $V_{eff}(z_{12})$ 
exhibits a Coulombic decay behavior at large electron 
separation. However, at small distances between the electrons this effective 
potential does not diverge at $z_1=z_2$ which is beneficial for numerical treatments:

\begin{eqnarray}
 \left.\frac{}{} V_{eff}(z_{12})\right|_{z_{12} \rightarrow
\infty}&\longrightarrow & 
\frac{1}{z_{12}}\left(1-\frac{l^2}{\left(z_{12}\right)^2}\right) 
\label{eff1_asymptotics} \\
\left. V_{eff}(z_{12})\right|_{z_{12} \rightarrow
0} &\longrightarrow & \frac{1}{l}\left(\sqrt{\frac{\pi}{2}}-\frac{z_{12}}{l}+ \cdots \right) 
\label{eff2_asymptotics}
\end{eqnarray}

The validity of the effective potential in different confinement 
regimes was studied in~[\onlinecite{bednarek_effective_2003}] for double QDs as a 
function of the distance $R$ between QDs. From 
Eq.~(\ref{eff1_asymptotics}) we see that
$l/z_{12}$ defines the correction order of the effective 
interaction at large distances. If we take 
the distance between the dots $R$ as a measure of the closest distance that 
electrons will be from each other, then $z_{12}/l\approx R/l$. We realize then 
from Eq.~(\ref{eff1_asymptotics}) that in the regime studied in this work 
($l\approx 1$ and $R\approx 10$), the electrons are already in the asymptotic 
regime of the effective potential. Notice also that the peak at $z_1=z_2$ 
scales as $1/l$ (see Eq.~(\ref{eff2_asymptotics})) indicating that in truly narrow
confinements ($l \rightarrow 0$) there is less room for the electrons to 
avoid the divergence of the Coulomb interaction.

\section[mctdh]{Computational Details}\label{methods}

The dynamical evolution of the system was obtained by solving the 
time-dependent electronic Schr\"odinger equation employing the 
multiconfiguration time-dependent Hartree
(MCTDH) approach.~\cite{meyer_multi-configurational_1990,
meyer_multidimensional_2009} The triplet wave function
\begin{equation}
\Psi(\ve{r}_1,\ve{r}_2,t) = \sum^n_{i,j} A_{ij}(t) \varphi_{i}(\ve{r}_1,t)
\varphi_{j}(\ve{r}_2,t),
\end{equation}

\noindent was expanded in time-dependent single particle functions
$\varphi_{i}(\ve{r},t)$~(SPFs) and coefficients $A_{ij}(t)$ that fulfill the
antisymmetry condition $A_{ij}(t)=- A_{ji}(t)$ for all times. 
The Dirac-Frenkel variational 
principle~\cite{dirac_note_1930,frenkel_wave_1934}
\begin{equation}
 \braket{\delta\Psi}{H-i\frac{\partial}{\partial t}}{\Psi}=0
\end{equation}

\noindent was used to obtain the equations of motion for the coefficients and 
SPFs.

They were efficiently solved using
a constant mean field approach as implemented in the MCTDH-Heidelberg 
package.~\cite{meyer_multidimensional_2009, beck_multiconfiguration_2000}
The convergence of numerical results was ensured by monitoring the population
of the least populated SPF. This is reasonable because the SPFs are 
adaptive in time and are optimized to describe $\Psi(\ve{r}_1,\ve{r}_2,t)$ with
the least possible number of SPFs.

The multimode SPFs $\varphi_i(\ve{r}_q,t)$ were in turn expanded in 
one-dimensional time-dependent SPFs for each 
of the Cartesian coordinates $(x,y,z)$ as
\begin{equation}
 \varphi_i(\ve{r}_q,t) = \sum_{lmn} C^{(q)}_{lmn}(t)\chi^{(x)}_l(x_q,t)
 \chi^{(y)}_m(y_q,t)\chi^{(z)}_n(z_q,t).
\end{equation}

These one-dimensional SPFs $\chi_l$ are expanded on a DVR-grid 
(discrete variable representation). 
 We chose harmonic 
oscillator DVRs for the $x$ and $y$, and a sine DVR
for the $z$ coordinate as listed in Tab. II.

In the full 3D calculations the Coulomb potential was regularized as
$1/r_{12} \rightarrow 1/\sqrt{r^2_{12}+a^2}$ with $a=0.01$
to prevent divergences at $\ve{r}_1=\ve{r}_2$, and then transformed into sums 
of products using the
POTFIT~\cite{beck_multiconfiguration_2000} algorithm of MCTDH.

A quadratic complex absorbing potential 
(CAP) was placed at the position 
$\pm z_{cap}$ along the $z$
coordinate to absorb the outgoing electron 
before it reaches the end of the DVR
grid. The CAP obeys
\begin{equation}\label{cap}
 W_{\pm}=-i\eta (z \mp z_{cap})^{2}\Theta(z \mp z_{cap})
\end{equation}
where $\eta$ is the CAP strength and $\Theta$ is the Heavyside step 
function.
The absorption prevents the unphysical reflection of outgoing electrons
at the grid boundaries. 

The absorption of the WP is also used to analyze the 
energy distribution of the outgoing WP. The quantity that we want to compute is the 
reaction probability (RP) for ICEC which corresponds to the scattering matrix element 
$|S_{L_0,R_0}(E_T)|^2$ which is the probability that an electron impinging 
from the left on the DQD possessing an electron bound at $R_0$ leads to emission 
of an electron to the right leaving behind a DQD with an electron bound to $L_0$. 

The computation of the matrix element was performed by using the expression for 
the stationary scattering eigenfunctions in terms of 
the initial wave packet $WP_i$~\cite{tannor_wave_1993} in order to obtain the 
amount of emitted density from the wave packet absorbed by the 
CAP.~\cite{beck_multiconfiguration_2000} The energy distribution $|\Delta_{WP_i}(E_T)|^2$ of the 
incoming WP$_i$ is used to normalize the Fourier transform of the absorbed 
density $g_{L_0}(\tau)$ to obtain the reaction probability 
(RP).~\cite{beck_multiconfiguration_2000} We explicitly computed 
\begin{equation}\label{scatt_matrix}
 \frac{RP(E_T)}{100}=|S_{L_0,R_0}(E_T)|^2 = \frac{2{\rm Re} 
 \int^{\infty}_0 g_{L_0}(\tau)e^{iE_Tt/\hbar} {\rm d} \tau}{\pi |\Delta_{WP_i}(E_T)|^2}
\end{equation}

\noindent where

\begin{eqnarray}\label{g}
 g_{L_0}(\tau) &=& \int^{\infty}_0 
 \braket{\Psi(t)}{P^{(1)}_{L_0}W^{(2)}_{+} 
P^{(1)}_{L_0}}{\Psi(t+\tau)}{\rm d}t \nonumber \\
 &+& \int^{\infty}_0 
 \braket{\Psi(t)}{P^{(2)}_{L_0}W^{(1)}_{+} 
P^{(2)}_{L_0}}{\Psi(t+\tau)}{\rm d}t \nonumber\\
 &=&\! 2\!\int^{\infty}_0\!\! 
\braket{\Psi(t)}{P^{(1)}_{L_0}W^{(2)}_{+} 
P^{(1)}_{L_0}}{\Psi(t+\tau)}{\rm d}t
\end{eqnarray}
\noindent and
\begin{equation}
 \Delta_{WP_i}(E_T) = \sqrt{\frac{m^*}{2\pi p_{R_0}}}\int^{\infty}_{-\infty} 
 f_{WP_i}(z)
 e^{ip_{R_0}z} {\rm d}z
\end{equation}

\noindent where the function $f_{WP_i}(z)$ is a Gaussian wave packet with a 
spatial width $\Delta x_{WP_i}$. $\Delta_{WP_i}(E_T)$ is the energy 
distribution of the incoming WP$_i$ peaked around
$\varepsilon_{WP_i}$ and given by the appropriate Fourier 
transform which uses the incoming momentum 
$p_{R_0}=\sqrt{2m^*(E_T-E_{R_0})}\equiv p_i$.~\cite{tannor_wave_1993}

$g_{L_0}(\tau)$ is the absorbed electronic density by the right CAP while 
another electron is bound in the $L_0$ state. The projectors $P^{(q)}_{L_0}$ 
acting on electron $q$ specify which electron is in the $L_0$ state, and the sum over
both possible configurations gives the total absorbed density. Note that this quantity explicitly correlates 
both events, emission and capture, and thus gives only the ICEC contribution of
the total emitted density. The scattering matrix in Eq.~(\ref{scatt_matrix}) 
corresponds to the $R_0$ initial state because the initial wave function 
\begin{eqnarray}
\Psi(0)&=&
\left[f_{WP_i}(z_1)\phi_{R_0}(z_2)- f_{WP_i}(z_2)\phi_{R_0}(z_1)\right]\times \nonumber\\ 
& &\phi_0(x_1,y_1)\phi_0(x_2,y_2) 
\end{eqnarray}
\noindent represents a bound electron at $R_0$ plus an incoming electron both in the 
ground state of the confinement potential.  

The RP is a wave-packet independent quantity in the energy range of the size 
of the energy width of the incoming wavepacket WP$_i$ 
(see Eq.~(\ref{scatt_matrix})). At each energy, the RP gives the relative 
amount (in \%) of the electron density that would be emitted in the calculation 
with a monoenergetic electron at that energy. The absorption of WP$_i$ by the 
CAP outside the DQD economizes the computation time needed to obtain
the RP.

\section[results]{Results}\label{results}

\begin{figure}
\includegraphics[width=0.25\textwidth]{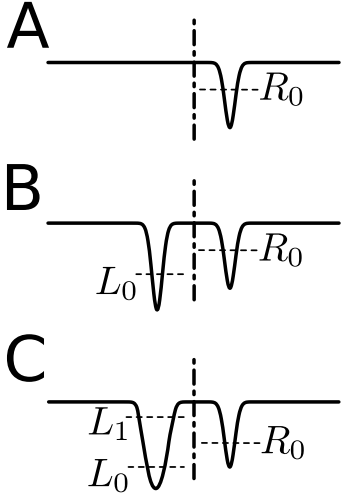}
\caption{\label{setups}The three QD model potential setups studied in this work.
Setup A is briefly analyzed and used only to clarify that no transmission to
the right is possible without a left QD. In setup B each QD has one bound 
one-electron state, $L_0$ and $R_0$, respectively, and B is used to show how ICEC 
works in double QDs.
In setup C the left QD has an additional one-electron excited state $L_1$. In such a
configuration the energy of the two-electron resonance $\ket{L_1R_0}$ can be
tuned to substantially increase the ICEC reaction probability.}
\end{figure}

In this section we analyze the electronic structure (Sec. V.A)
and the dynamics of the electrons (Sec. V.B) in the DQD relevant for ICEC. 
We compare a number of different configurations that can be classified 
according to the general setups of the QD model potentials shown in 
Fig.~\ref{setups}. In setup A only the right QD with a single one-electron state
$R_0$ is present. The only purpose
of investigating this setup is to prove that, for the incoming 
electron energies considered in this work, no transmission 
to the right is 
possible when the left QD is not present. The configurations belonging to 
setup B have one left and one right QD and each dot has a single one-electron 
state, $L_0$ and $R_0$, respectively. In these cases ICEC is allowed~\cite{pont_controlled_2013} and occurs 
as visualized in Fig.~\ref{scheme}.
Finally, setup C comprises configurations where the left QD has an
excited one-electron state $L_1$ in addition to the $L_0$ ground state allowing 
for the intermediate state $\ket{L_1R_0}$ to be formed. Since
electrons located in the left and right QD are interacting with each other 
through the long-range Coulomb interaction pushing the state into the continuum, 
this state turns out to be a \emph{two-electron} resonance. We will show that 
under certain conditions this resonance leads to a remarkable increase of 
the ICEC probability.

\subsection[elec_struc]{Electronic Structure}\label{elec_struc}

As a first step in our analysis we want to study the electronic structure of
the DQD embedded in the wire. As explained in Sec.~\ref{model} the
two-electron states can be named after the one-electron states of the DQD.
The confinement part of the wave function is
described by the lowest energy harmonic oscillator wave functions in $x$ and $y$ both with 
frequency $\omega$ and 
effective mass $m^*$ and we therefore concentrate 
only on the $z$ wave function analysis in what follows.

\begin{figure}[t]
\includegraphics[width=0.45\textwidth]{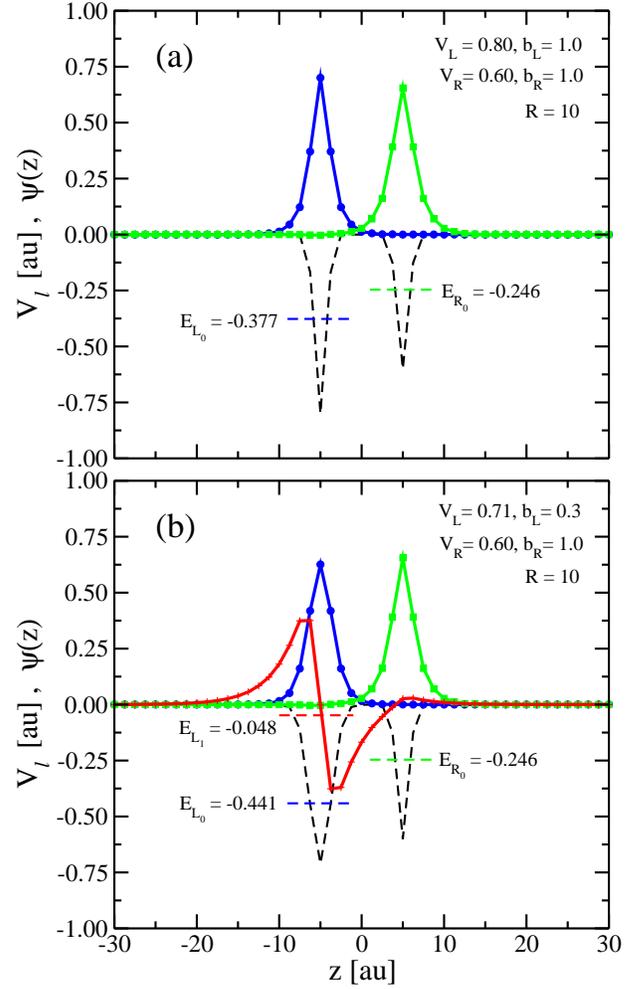}
\caption{\label{pes_dots_states}(Color online) The potential $V_{l}(z)$ 
(black dashed lines) and its bound states for two different configurations. 
(a) The DQD potential binds only two one-electron states with wave functions 
$\psi_{R_0}$ (green squares) and $\psi_{L_0}$ (blue circles). The energy levels of 
these states are marked by dashed lines and the respective binding
energies are indicated. (b) The left well is shallower and wider than in (a) 
and binds one additional excited one-electron state with a p-type wave 
function $\psi_{L_{1}}$ (red crosses).}
\end{figure}

The potential energy curves and the wave functions of the states for two of the 
configurations used in the dynamical calculations are shown in 
Fig.~\ref{pes_dots_states}. The concrete configuration of setup B in 
Fig.~\ref{pes_dots_states}(a) has two bound one-electron states
$L_0$ and $R_0$. It is clearly visible that both states are localized in the respective QDs and 
that there is no hybridization of the states. Two characteristics of this 
configuration make this possible. One is the distance $R$ between the
QDs, which is large
compared to their size, and the other is the asymmetry of the DQD which
leads to different energies for the left and right QDs. 

The configuration shown in
Fig.~\ref{pes_dots_states}(b) is a representative of setup C. It shows a wider
and shallower left QD which allows
for an excited one-electron state $L_1$. 
We see that the binding energy $E_{L_1}$ is much smaller than $E_{L_0}$ and $E_{R_0}$
and the wave function $\psi_{L_1}$ is therefore more extended than 
$\psi_{L_0}$ and $\psi_{R_0}$. 

We set the origin of the energy scale to
$2\omega$ throughout the study. 
It amounts to the energy contributed 
by both electrons in the ground state of the transversal confinement potential
$V_{c}$ (Eq.~(\ref{vc})). 
With this choice the bound (unbound) states of the longitudinal potential of the DQD 
have negative (positive) energies.

\subsection[dyn_calc]{Dynamical calculations and results}\label{dyncalc}

By employing electron dynamics calculations we can investigate what happens 
when an electron coming from the left side approaches the DQD where 
one electron is initially bound and how, if at all, ICEC occurs. We start 
with the simplest case of setup A (Sec. \ref{singledot}) where only the
right QD is present and then move on to different configurations
of setup B (Sec. \ref{icecdqd}) and C (Sec. \ref{resicec}). 
All examples were computed using both the 1D model (Sec.~\ref{1dmodel}) and the full
3D Hamiltonian (Sec.~\ref{hamiltonian}) for triplet symmetry. In all cases we 
chose the energy of the incoming wave packet (WP$_{i}$) such that it is to 
low to ionize the electron initially bound to the $R_0$ state, even if the full energy width of the
WP$_i$ is considered. 

\subsubsection[singleQD]{One single QD}\label{singledot}

The initial state of the two-electron systems is an incoming free electron
from the left and a bound one in the right QD. A similar setup was studied
before,~\cite{selsto_absorbing_2010} however, for a different energy regime of the
incoming electron in which two-electron ionization was allowed. The parameters
$V_{R}=0.6$~a.u. and $b_{R}=1.0$~a.u. used here give a single bound
state with an energy of $E_{R_0}=~\!-0.2463$~a.u. The incoming wave packet
(WP$_{i}$) is an energy normalized Gaussian peaked around
$\varepsilon_{WP_i}=0.056$~a.u. The packet has a spatial width 
$\Delta x_{WP_i}=10.0$~a.u. and an energy width 
$\Delta \varepsilon_{WP_i}\approx 0.033$ a.u.~\footnote{The width
of the Gaussian wave packet in momentum space is given by $\Delta p=\frac{1}{2\Delta x}$. Then the energy width is given by 
$\Delta \varepsilon_{WP_i}=p_{i}\Delta p = \frac{\sqrt{2\varepsilon_{i}}}{2 \Delta x}$}
which is not enough to ionize the bound electron by the incoming one.  
Moreover, excitation to higher states in
the transversal directions are energetically forbidden for these parameters.

\begin{table*}
\centering 
\caption{Parameters used in the MCTDH calculations. The discrete variable 
representation (DVR) types correspond to harmonic oscillator (HO) and sine DVR 
(SIN).}
\label{tab:mctdh_par} 
\begin{ruledtabular}
\begin{tabular}{l c c c } 
& $x$ & $y$ & $z$ \\
\hline
DVR type& HO & HO & SIN \\ 
DVR points/Primitive Basis & $5$ & $5$ & $431$ \\
Range / a.u. & $(-2.02,2.02)$ & $(-2.02,2.02)$ & $(-270.00,270.00)$ \\ 
Grid Spacing (d$x$) / a.u. & $1.01$ & $1.01$ & $1.25$ \\ 
SPFs &  & $14$ ($x,y,z$ combined) & \\ 
$z_{cap}$ & - & - & $\ 168.75$\\ 
\end{tabular}
\end{ruledtabular}
\end{table*}

\begin{figure}
\includegraphics[width=0.48\textwidth]{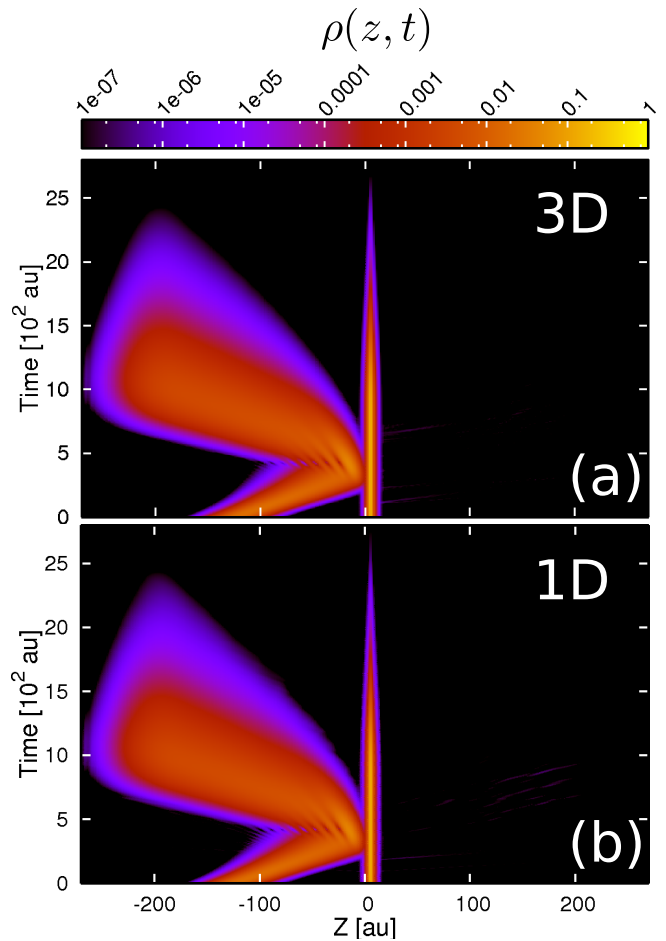}
\caption{\label{surface_plots_A} (Color online) 
Evolution of the electronic density Eq.(17) for a QD of setup A using the full 
three-dimensional Hamiltonian (a) and (b) the one-dimensional model of section~\ref{1dmodel}. 
The incoming wave packet (WP$_i$) approaches from the
left to the QD located at $z=5$ a.u. (right dot) which is initially occupied by
one electron ($R_0$ state). The incoming packet is initially located at
$z=-125$ a.u. with energy $\varepsilon_{WP_{i}}=0.056$ a.u. and has a spatial 
width $\Delta x_{WP_i}=10.0$ a.u. and energy width 
$\Delta \varepsilon_{WP_{i}}=0.033$ a.u. The parameters used for the 
MCTDH simulations are given in Tab.~\ref{tab:mctdh_par}. Note that the energy covered
by the $WP_i$ is to low to remove the electron in the right QD (
$\varepsilon_{i} + \Delta \varepsilon_{WP_{i}} < |E_{R_{0}}|$). 
Since the left QD is missing, no emission to the right is observed.}
\end{figure}

The dynamics of the full 3D scattering process calculated according to 
the method described in Sec.~\ref{methods} is visualized
in Fig.~\ref{surface_plots_A}(a) by the longitudinal electronic density
\begin{equation}
\rho(z,t)=\int \textrm{d}\ve{r'} \int\textrm{d}x \int\textrm{d}y
\left|\Psi(\ve{r},\ve{r'},t)\right|^2 
\end{equation}
as a function of $z$ and $t$. The incoming electron is
completely reflected starting at about $t$=3 a.u. while the other electron 
remains bound in the right QD. The same calculation was made using the 
one-dimensional model described in Sec.~\ref{1dmodel} and is shown in 
Fig.~\ref{surface_plots_A}(b) for comparison. 
The evolution is in both cases very similar, only the population $P$ of the 
lowest populated SPF (which is a measure of the convergence as explained in 
Sec.~\ref{methods})
is different (but however small) in each case giving a value of 
$P=1 \times 10^{-8}$ for the simplified model and $P=1 \times
10^{-7}$ for the full calculation.
For long times ($t\approx 25$ a.u.) the total density 
$\rho(z,t)$ in the system decreases to zero. The reason for this unphysical 
behavior is the CAP absorbing the continuum electron. This effect has no 
impact on the observed results, because the reflection process 
is already completed within a much shorter time of about 10 a.u.

\subsubsection[icecdqd]{ICEC in a double quantum dot}\label{icecdqd}

\begin{figure*}
\includegraphics[width=0.9\textwidth]{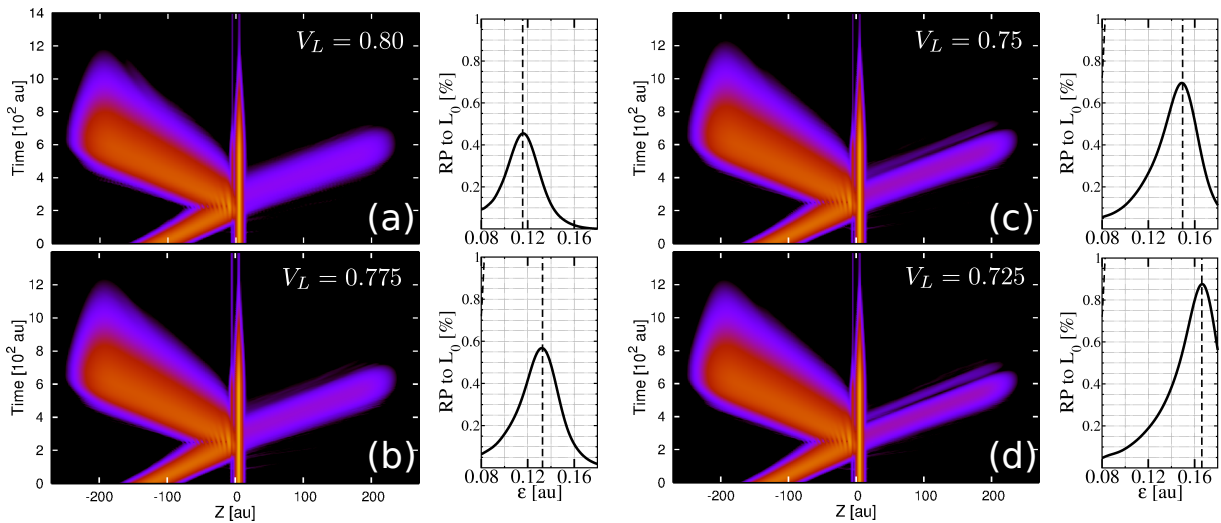}
\caption{\label{surface_plots_B}(Color online) 
Evolution of the electronic density (left panels) and the obtained ICEC 
reaction probabilities (RP) (right panels) for setup B of Fig.~\ref{setups}. 
The incoming wave packet (WP$_i$) approaches the DQD centered at $z=0$ a.u. 
from the left which is initially occupied by
an electron in the right QD ($R_0$ state).
$\varepsilon_{i}=0.130$ a.u. and the 
parameters of the right potential are the same in all four cases. The 
depth of the left dot $V_L$ is varied as indicated for each case: (a) $V_L=0.800$,
(b) $V_L=0.775$, (c) $V_L=0.750$, 
(d) $V_L=0.725$ a.u. The emission of the electron seen to the right is from $R_0$ 
and takes place through ICEC (compare Fig.~\ref{scheme}). The emitted packet 
acquires momentum in all cases according to the energy conservation (Eq.~(\ref{energy_cons})) 
and is faster than the incoming packet. This is clearly visible from 
the slope of the outgoing density which is smaller than that of the incoming 
density. The reaction probabilities shown in the right panels exhibit a peaked 
energy distribution centered at the values $\varepsilon_i^{(peak)}$ (depicted 
as dashed vertical lines computed by Eq.~(\ref{e_peak}) and listed in Tab.~\ref{tab:energ_setup_B}).}
\end{figure*}

We now focus on configurations of setup B where we added the left QD at a
distance $R = 10.0$ a.u. ICEC takes place in these DQDs as depicted in the
scheme in Fig.~\ref{scheme} and we confirm this by using different configurations
for which Eq.~(\ref{energy_cons}) is shown to be fulfilled. The spatially resolved
time evolution of $\rho(z,t)$ of four configurations is shown in each left panel of
Fig.~\ref{surface_plots_B} (a)-(d).
The right QD and the incoming wave packet WP$_i$ are the same in all four
configurations with $V_R=0.6$ a.u., $b_R = 1.0$ a.u. (same as for setup A before) and 
$\varepsilon_i=0.130$~a.u., $\Delta x_{WP_i}=10$ a.u., 
$\Delta \varepsilon_{WP_i}=0.051$ a.u. The left QD is characterized by 
$b_L=1.0$~a.u., but its depth varies in these configurations taking on the values
$V_L=0.800,0.775,0.750,0.725$ a.u. The corresponding
energies $E_{L_0}$ and $\Delta E = E_{R_0}-E_{L_0}$ are given in Tab.~\ref{tab:energ_setup_B}.

\begin{table} 
\centering 
\caption{
The parameters used in the four configurations for setup B discussed in the 
text and in Fig.~\ref{surface_plots_B}, and the resulting computed energies,
final momenta $p_{f}$, and positions $\varepsilon^{(peak)}_{i}$ of the peak 
values of the reaction probability (RP). All values are given in a.u.}
\begin{ruledtabular}
\begin{tabular}{l c c c c} 
$V_L$ & $E_{L_0}$  & $\Delta E$ & $p_f$ & 
$\varepsilon^{(peak)}_i$ \\ 
\hline
$0.800$ & $-0.3769$  & $0.1306$ & $0.722$  & $0.1157$ \\
$0.775$ & $-0.3599$ & $0.1136$ & $0.698$ & $0.1327$ \\ 
$0.750$ & $-0.3430$  & $0.0967$ & $0.673$ & $0.1496$ \\ 
$0.725$ & $-0.3264$ & $0.0801$ & $0.648$ & $0.1662$ \\
\end{tabular}
\end{ruledtabular}
\label{tab:energ_setup_B} 
\end{table}

Electron emission to the right is clearly visible in all four cases. The flatter
slope of the final wave packet (WP$_{f}$) trajectory traveling to the right
indicates that the emitted electron has higher momentum than the incoming 
electron. According to Eq.~(\ref{energy_cons}) the
final energy of the outgoing electron represented by $p_f$ calculated from 
Eq.~(\ref{final_momentum}) (see Tab. \ref{tab:energ_setup_B}) decreases 
when the depth $V_L$ decreases. 
The RP gives a quantitative measure of ICEC 
and can be computed {using Eq.~(\ref{scatt_matrix})}. Descriptively, 
it is the probability of capturing an electron in the left QD while simultaneously 
emitting an electron to the right from the right QD. The RP
as a function of the incoming electron energy $\varepsilon_i$ is shown in each 
right panel of Fig.~\ref{surface_plots_B}~(a)-(d).
The energy range covered in the RP plots is determined by the peak 
$\varepsilon_{WP_i}$ 
with the energy width $\Delta \varepsilon_{WP_i}$ 
of the incoming wave packet. It is possible 
to obtain reliable results from one simulation within the energy range 
$\varepsilon_{WP_i}\pm 2 \Delta \varepsilon_{WP_i}$, which is used for the RP plots.

At this point we would like to discuss more the meaning of the RP. 
The values given in the plots for ICEC are exactly the amount of the total 
electron density in percent that would be ejected from $R_0$ to the 
right and correspondingly the increase of the population of $L_0$, 
if the electron incoming from the left was mono-energetic with energy 
$\varepsilon_i$. On the other hand, a mono-energetic electron implies 
an infinitely wide WP$_i$ ($\Delta x_{WP_i} \rightarrow \infty$ ), which cannot 
be realized numerically on our finite DVR grid. 
In our calculations we take a rather broad incoming wavepacket and by 
employing Eq. (19) we can compute the RP.

Let us analyze the results for RP shown in 
Fig.~\ref{surface_plots_B}. They clearly show that ICEC is no at all constant 
or 
even monotonic in the covered energy range. On the contrary, it is seen that ICEC is very 
selective in energy. This is a non-trivial result considering that the ICEC 
channel into $L_0$ is open for all incoming electron energies 
(Eq.~(\ref{final_momentum})). The peak of the RP has its origin in the fact that the 
total energy $E_T$ (see Eqs.~(\ref{total_energy_in}) and (\ref{total_energy_out})) is the 
relevant energy in a scattering process.~\cite{taylor_scattering_2006} The RP 
shows a marked increase in the probability when the total energy $E_T$ matches 
the energy gained by the DQD~($-\Delta E$) in the ICEC 
process in which the emitted 
electron takes an energy $\Delta E$. Using Eq.~(\ref{total_energy_in}) 
we obtain the value of $\varepsilon_i$ at which the peak of 
the RP is located,
\begin{equation}\label{e_peak}
\varepsilon^{(peak)}_i = -E_{R_0} - \Delta E.
\end{equation}

The values obtained for $\varepsilon^{(peak)}_i$ are given in Tab.~\ref{tab:energ_setup_B} and
depicted with vertical dashed lines in the RP plots of Fig.~\ref{surface_plots_B}. We see
that the RP peaks obtained from the dynamics fit exactly the predicted values 
using Eq.~(\ref{e_peak}). The RP values for the configurations of setup B all revealed probabilities 
below $1\%$.

\subsubsection[resicec]{Capture in the presence 
of a two-electron resonance}\label{resicec}

The physics of the capture is complicated in the presence of an increased 
number of bound states of the QDs. In general, several capture 
and decay channels will be open before and after the capture and the physics of
resonance states comes into play. We analyze the probably most simple extension to the
DQDs described in the previous sections (setups A and B) by including one extra
excited state in the left QD (setup C).

\begin{figure}
\includegraphics[width=0.5\textwidth]{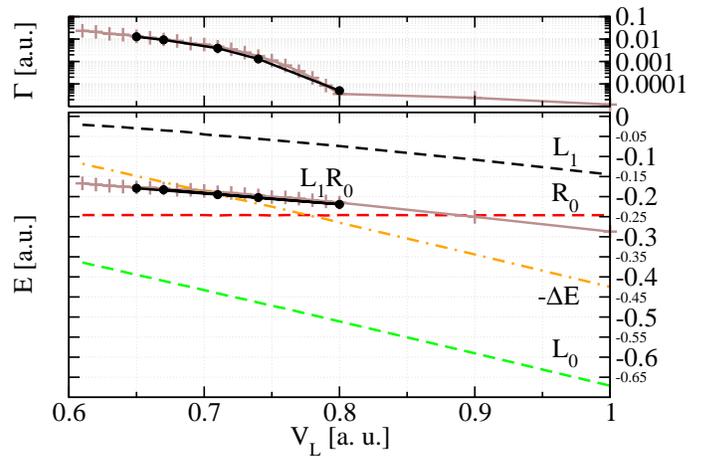}
\caption{\label{E_and_G_vs_vl}(Color online) Width (top panel) and energies 
(bottom panel) of the $\ket{L_{1}R_{0}}$ two-electron resonance playing a 
relevant role in enhancing the ICEC probability in setup C as function of the 
depth $V_L$ of the left QD. The energy and width of the resonance obtained in the 
one-dimensional model of section~\ref{1dmodel} are shown with brown crosses 
and those obtained for the full three-dimensional system with black dots. Both 
sets of results are very similar. Shown are also the energies of all the 
single-electron states computed for the full three-dimensional system. The 
energies of the $L_{0}$, $R_{0}$, and $L_{1}$ states of the DQD are depicted 
as dashed green, red, and black lines, respectively, while the value of 
the energy difference $-\Delta E= E_{L_0}-E_{R_0}$ is indicated by 
a dash-dotted 
orange line.}
\end{figure}

Accordingly, we modify the potential well of the left QD by choosing $b_L=0.3$
a.u. instead of $b_L=1.0$ a.u., i.e. we make the left well wider. Then we 
analyze the energies of the states as a function of the depth $V_L$. 
This dependence is shown in Fig.~\ref{E_and_G_vs_vl} for the three-dimensional model.
Due to the Coulomb interaction the DQD 
accommodates a two-electron resonance which derives from the one-electron
states $L_1$ and $R_0$. The $\ket{L_1R_0}$ resonance energy
and decay rate (inverse lifetime) are shown as black dots in
Fig.~\ref{E_and_G_vs_vl}. Decay rates in QDs can be computed using different
methods.~\cite{bande_dynamics_2011, cherkes_electron_2011, pont_entropy_2010}
We follow here the approach employed in~\cite{bande_dynamics_2011} in which 
the resonance state $\ket{L_1R_0}$ is prepared by imaginary
time propagation followed by the real time evolution to find its total decay rate.

\begin{figure}
\includegraphics[width=0.45\textwidth]{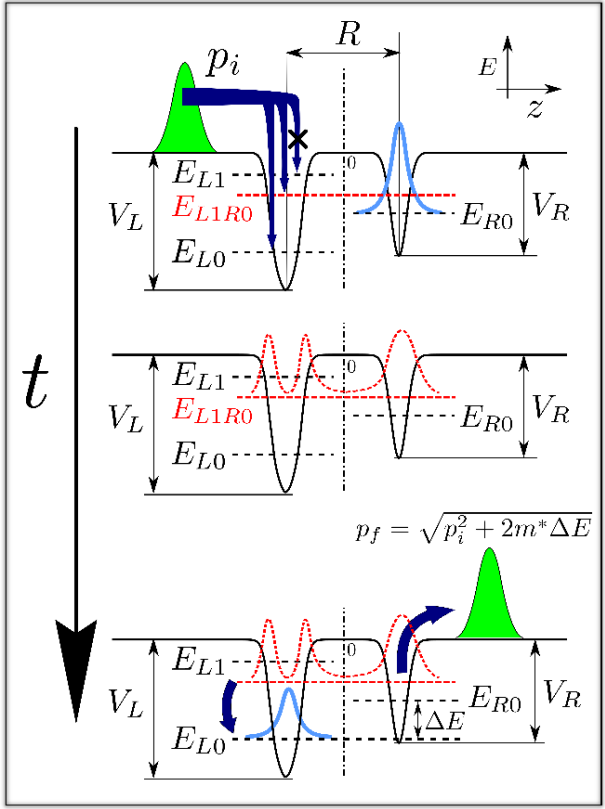}
\caption{\label{scheme_res}(Color online) Schematic view of
interatomic Coulombic electron capture in a model potential for a double QD 
in the presence of a two-electron resonance $\ket{L_1R_0}$ (dashed red lines).
The incoming electron can be captured into $\ket{L_1R_0}$ (middle panel) 
because the resonance energy lies above the threshold. Then, the 
resonance decays by ICD (middle to bottom panel), 
a process in which the excited electron of the left QD decays from $\ket{L_{1}}$ 
to the $\ket{L_0}$ state while transferring the excess energy to the electron 
in the right QD which is emitted to the continuum.}
\end{figure}

 The capture process occurs in the presence of the resonance as
indicated in Fig.~\ref{scheme_res} so that different electron capture scenarios can be imagined.

As before in setup B, electron capture into the $L_0$ state with 
simultaneous release of the other electron from the $R_0$ state 
is one possible pathway (direct ICEC). Moreover, if the energy of the resonance is
above the threshold, the incoming electron can be captured into the 
two-electron resonance state $\ket{L_1R_0}$. After this it decays through a
process called interatomic Coulombic decay (ICD),~\cite{bande_dynamics_2011,
cherkes_electron_2011, cederbaum_giant_1997, sisourat_interatomic_2010,
sisourat_ultralong-range_2010, jahnke_ultrafast_2010} that means by 
deexcitation of the electron in the left QD ($\ket{L_1} \rightarrow \ket{L_0}$). 
The released energy is used to emit the
electron from the right QD ($\ket{R_0}\rightarrow
e^-$).~\cite{bande_dynamics_2011} We denote this pathway as the resonance 
channel and the process as resonance-enhanced ICEC. After being populated 
by the incoming electron, the resonance can 
also decay by emitting elastically the electron to the left. This decay 
resembles that of a shape resonance:~\cite{taylor_scattering_2006}
$e^{-} + \ket{R_0}\rightarrow \ket{L_1R_0} \rightarrow \ket{R_0} + e^-$.  This decay
is of course only possible when the resonance energy $E_{L_1R_0}$ is higher 
than $E_{R_0}$, a situation that was not usually fulfilled in the systems where
ICD was investigated earlier. For completeness we mention that the 
incoming electron energy is sufficiently low so that direct electron capture 
into the $L_1$ state is energetically forbidden for all cases considered here.

\begin{figure*}
\includegraphics[width=0.9\textwidth]{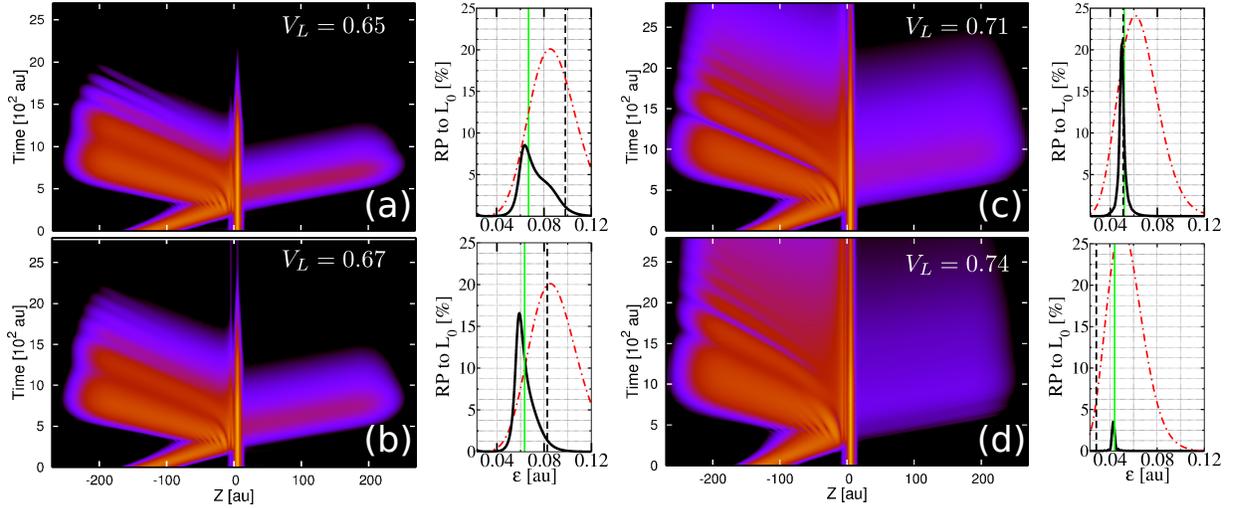}
\caption{\label{surface_plots_C}(Color online) Evolution of the electronic 
density and obtained reaction probability (RP) for setup C of Fig.~\ref{setups}.
The incoming wave packet (WP$_i$) approaches from the
left to the DQD centered at $z=0$ a.u. which is initially occupied by
an electron in the right QD ($R_0$ state). The left dot binds two states $L_0$ and $L_1$
and the depth of the left dot $V_L$ is varied as:
(a) $V_L=0.650$, (b) $V_L=0.670$, (c) $V_L=0.710$, (d) $V_L=0.740$. The emission of 
the electron initially located in $R_0$ takes place through the process shown in
Fig.~\ref{scheme_res}. In the right panels the energy of the two electron resonance 
($L_1R_0$) and that of the direct ICEC peak are indicated with vertical lines (green
continuous and black dashed, respectively) and tabulated in Tab.~\ref{tab:energ_setup_C}. 
For the simulations we used different WP$_{i}$ (red dash-dotted line), because 
the resonance and direct ICEC peak energies vary and then the relevant region of 
the RP is different for each case. The RP (black solid line) shows 
two distinguishable contributions to the energy distribution in case (a): 
one from the resonance state (peak, 0.060 a.u.) and the other 
from direct ICEC (shoulder, 0.085 a.u.). For case (b) both contributions have
nearly the same energy and the emission increased markedly near the resonance
energy. In case (c) the matching of both energies (resonance and direct) 
gives a huge enhancement of the emission with a narrow energy distribution, which
corresponds to the width of the resonance (see table~\ref{tab:energ_setup_C}). 
The enhancement is lost in case (d) where the energy mismatch between the 
resonance and direct ICEC is enough to destroy the correlation of the processes.}
\end{figure*}

The time evolution of the electron density $\rho(z,t)$ has been 
calculated for different left well depths $V_L=$0.65, 0.67, 0.71, and 0.74 a.u. 
(Fig.~\ref{surface_plots_C}, left panels). Comparing with the results for 
setup B (Fig.~\ref{surface_plots_B}) a clear difference is observed 
for the density emitted from $z=0$ to the right. In setups $C$ a
continuous decay with an exponential time constant is visible while an almost 
instantaneous electron emission takes place for setups $B$. This indicates that
the mechanisms involved in the capture and emission processes are 
different for both setups. It is also noteworthy that the emitted electronic density to the left becomes more complex 
in case $C$ showing clear signatures of interference with the incoming WP$_i$. The
electron emitted elastically to the left is responsible for these interference effects. 

The results obtained for ICEC in section~\ref{icecdqd} show that the ICEC
probability is highest if the total energy $E_T$ matches the negative of
the energy difference $\Delta E$. It is, therefore, worthwhile to study the behavior
of the ICEC probability in relation to the value
of $\Delta E$ in the presence of a resonance. Fig.~\ref{E_and_G_vs_vl} shows that the resonance energy
crosses $-\Delta E$ around the value $V_L=0.70$ a.u. We
previously addressed the configuration with
$V_L=0.71$ a.u. which is near the crossing point of the
energies $E_{L_1R_0}=-\Delta E$.~\cite{pont_controlled_2013} In this case, 
the coincidence of the RP peak and the 
resonance energy lead to an extraordinary
increase of the ICEC probability. The presence of the resonance
enables an extra channel that can be tuned to cooperatively augment the emission.
The RP for this and three other $V_L$ values belonging to configurations 
above and below the mentioned crossing point are shown in the right panels of 
Fig.~\ref{surface_plots_C}. The incoming
WP$_i$ also depicted in Fig.~\ref{surface_plots_C} is different for each of the
configurations because the RP region of interest changes with the resonance
energy. Nevertheless, the energy range shown is the same in the four plots. 

We observe that for $V_L=0.65$ and $0.67$ a.u. the RP develops one large peak 
with a shoulder indicating a second peak. These two peaks correspond to 
the direct and the resonance-enhanced ICEC channels of the scattering process. 
The vertical lines depicted in the corresponding panels of 
Fig.~\ref{surface_plots_C} stand for the energy of the resonance and of the 
ICEC peak computed from Eq.~(\ref{e_peak}). The maxima of the RP are seen to be 
slightly displaced from these lines. In this sense the simple picture of 
independent resonance and direct ICEC peaks is not strictly valid and a 
correction taking the interaction between them into account is needed in 
order to obtain the correct peak positions. It should also be clear that both 
channels may interfere. It is noteworthy that the RPs now take on values of 
10 and 16 \%, respectively, which are substantially higher
than in the case of setup B where only the direct ICEC channel is 
operative.

{The choice of $V_L=0.71$ a.u. in panel (c) provides an 
extraordinary 
increase of the capture and emission probability. This probability of 22 \% 
indicates that the direct and resonance ICEC pathways coherently contribute 
to the same channel $\ket{R_0} + e^{-}$. The peak height strongly 
depends on whether the values of $E_{res}$ and $-\Delta E$ 
(depicted in Fig.~\ref{surface_plots_C} and listed in 
Tab.~\ref{tab:energ_setup_C}) coincide. We see in Fig.~\ref{surface_plots_C} for 
case (d) where $V_{L}$ is slightly enhanced that the peak height, 
now about 5 \%, is again smaller than in case (c). Clearly, the
increase of the ICEC probability in case (c) derives from the concurrence of 
both processes. The }{total width of the RP peak for case (c) is 
very narrow 
and given by the inverse lifetime of the resonance, as opposed to the other 
cases where a wider RP with more than one peak is obtained. This narrowness 
can be utilized to design an energy selective 
device.~\cite{pont_controlled_2013}
}

In case (c) the emitted electron density reaches the grid boundary before the resonant emission from
the DQD has terminated. This has no effect on the RP values as we find when using longer grids where
the full emission is possible before reaching the absorbing boundary. This is demonstrated explicitly in
the following section.
\begin{table} 
\centering 
\caption{Depth $V_{L}$ of the left QD, resonance and ICEC peak values in a.u. for the setup C cases.}
\begin{ruledtabular}
\begin{tabular}{l c c c c c } 
$V_L$ & $E_{res}$ & $-\Delta E$ & $\varepsilon^{(peak)}_i$ & $\Gamma${\tiny$(\times 10^{-4})$} & $\Gamma^{\textrm{\tiny(RP)}}${\tiny$(\times 10^{-4})$} \\ 
\hline
$ 0.65$ & $-0.179\pm0.003 $ & $-0.148$ & $0.0980$ & $130\pm9$ & $130\pm10$ \\
$ 0.67$ & $-0.183\pm0.008 $ & $-0.164$ & $0.0826$ & $92\pm6$ & $95\pm8$ \\
$ 0.71$ & $-0.196\pm0.002 $ & $-0.194$ & $0.0518$ & $39\pm2$ & $38\pm2$ \\
$ 0.74$ & $-0.202\pm0.002 $ & $-0.218$ & $0.0285$ & $26\pm2$ & $23\pm4$ 
\end{tabular}
\end{ruledtabular}
\label{tab:energ_setup_C} 
\end{table}

\subsubsection[1dicec]{ICEC in the one-dimensional effective model}\label{1dicec}

In addition to the results given by the full three-dimensional
simulations we performed computations using the one-dimensional model described
in section~\ref{1dmodel}. These calculations are much less time consuming and also 
allow to use much larger grids. 

\begin{figure}
\includegraphics[width=0.45\textwidth]{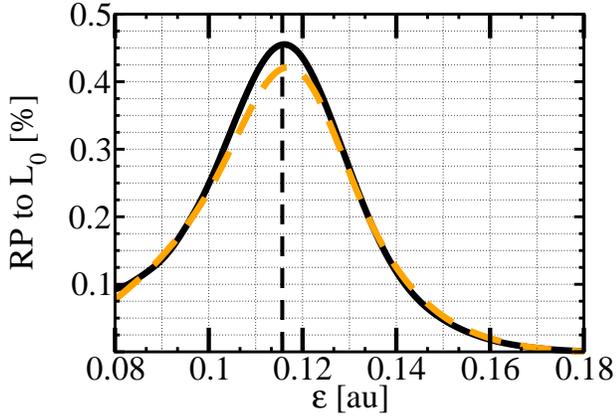}
\caption{\label{surface_plots_B1d}(Color online) Comparison of the ICEC reaction
probability (RP) for configuration (a) of setup B (see Fig. ~\ref{surface_plots_B}) obtained
using the one-dimensional effective model of Sec.~\ref{1dmodel} and the result of 
the full 3D computation. The RP of the full calculation (black) compares 
very well to the one-dimensional result (orange dashed). The
vertical black dashed line indicates the value of $\varepsilon_i^{(peak)}$ 
given by Eq.~(\ref{e_peak}).}
\end{figure}

The result for configuration (a) of Setup B is shown in 
Fig.~\ref{surface_plots_B1d} demonstrating that the RP is structurally
and quantitatively similar to that of the full three-dimensional computation. 
Without showing the picture we note that also the evolution of the electron 
density in the one-dimensional effective model is very similar to that of 
Fig.~\ref{surface_plots_B} for the full three-dimensional computation.

\begin{figure}
\includegraphics[width=0.45\textwidth]{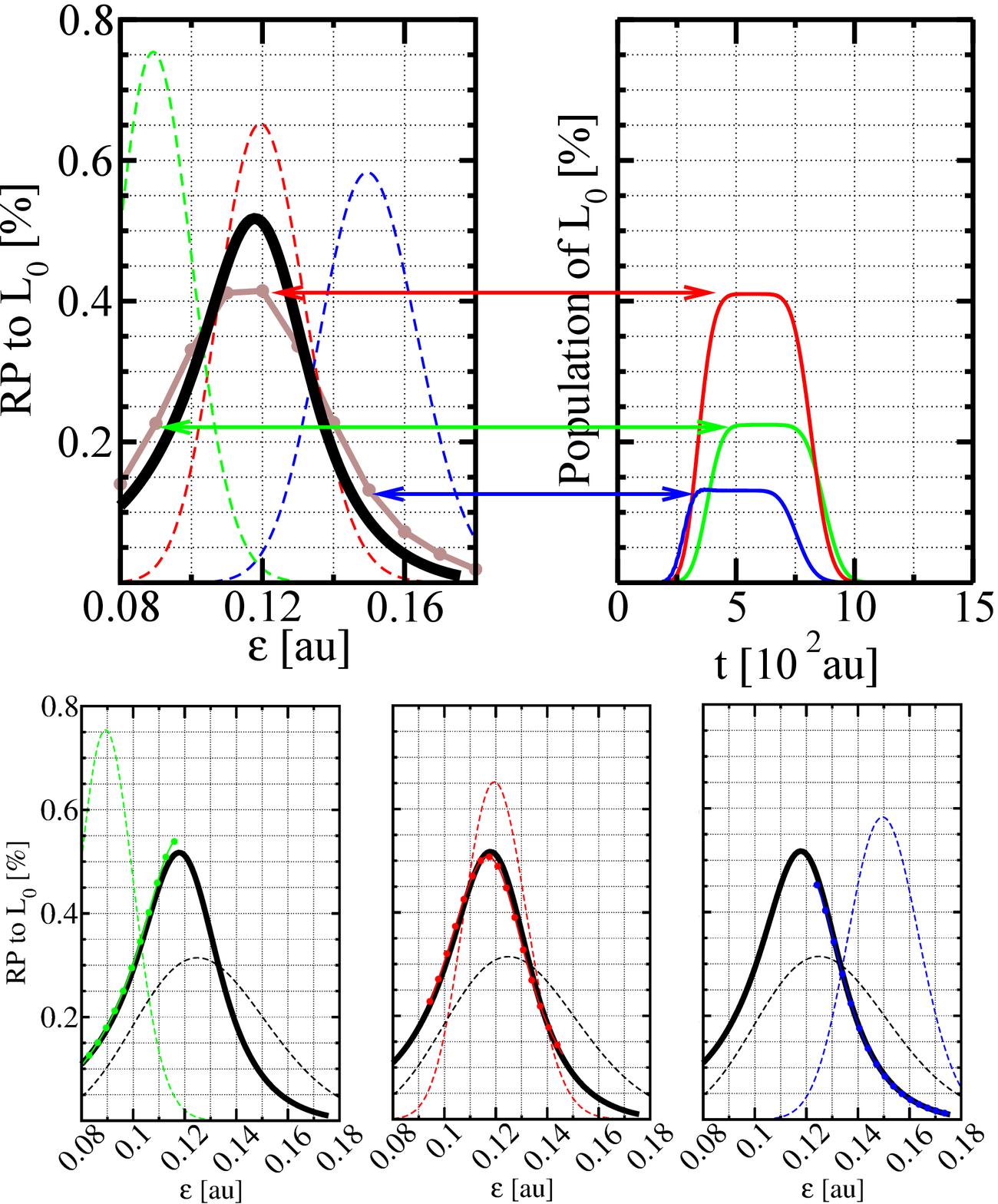}
\caption{\label{pop_1d}(Color online) Upper panels: Comparison of 
two methods to determine 
the reaction probability RP for case (a) of setup B. The one-dimensional model 
was employed. The left panel shows the RP obtained by the flux analysis using 
Eq.~(\ref{scatt_matrix}) with a single energetically wide WP$_i$,  
$\varepsilon_{WP_i}=0.130, \Delta \varepsilon_{WP_{i}}=0.0255$ a.u. and 
$\Delta x_{WP_i}=10.0$ a.u. (black line) and that obtained from several 
simulations at individual energies at the maxima of $L_{0}$ population of 
energetically narrower WP$_i$'s, 
$\Delta x_{WP_i}=20.0$, $\Delta \varepsilon_{WP_{i}}\approx0.0130$ a.u. 
(brown line). Regarding the latter, the population of $L_{0}$ as a function of 
time for three different WP$_i$'s 
$\varepsilon_{WP_i}= 0.090$, $0.120$, $0.150$ a.u. (green middle, 
red top and blue bottom line, respectively) is shown in the right panel. The
corresponding WP$_i$'s are shown in the left panel 
(dashed lines, green on the left, red in the middle and blue in the right). 
To obtain the brown curve in the left panel each maximum of the $L_{0}$ population 
was assigned to the respective $\varepsilon_{WP_i}$ in the left panel and values 
were interpolated.
Lower panels: Comparison of the reaction probabilities (RPs) 
obtained from 
different WPs. The RP from the wide WP$_i$ shown in the upper 
panels (black solid line) is compared with the RP obtained from  
the energetically narrower WP$_i$s, 
$\varepsilon_{WP_i}=0.090$, $0.120$, $0.150$ a.u., $\Delta \varepsilon_{WP_{i}}\approx0.0130$ 
a.u., $\Delta x_{WP_i}=20.0$ a.u. (green, red, and blue solid line with full 
circles). The corresponding WP$_i$s are shown with dashed lines.} 
\end{figure}

Since the computation times are considerably reduced for the one-dimensional 
model, we can perform the simulations on much longer grids than those 
used for the 3D calculations. Now, we can address numerically the question 
whether the RP obtained from Eq.~(\ref{scatt_matrix}) reproduces the population 
of the $L_{0}$ state via ICEC computed by employing incoming mono-energetic 
electrons.  The initial wave packet WP$_i$ can now be chosen to be spatially 
wider with $\Delta x=20$ a.u., with a reduced dispersion in energy 
$\Delta \varepsilon_{WP_{i}}\approx 0.0130$ a.u. As indicated in 
Sec.~\ref{resicec}, the maximum population of the $L_0$ state over time can now 
be computed for a selected value of the energy $\varepsilon_{i}$ 
of the incoming electron. This determines the RP at that energy. 
Clearly, we need to repeat the simulation using different incoming energies 
in order to construct a full RP curve. An example of an RP curve constructed 
in this manner is depicted in Fig.~\ref{pop_1d}. We observe that the maxima of
the $L_{0}$ populations follow closely the values of the RPs obtained from the 
flux determined via Eq.~(\ref{scatt_matrix}), even though the energy 
distributions $\Delta \varepsilon_{WP_{i}}$ of the WP$_i$s used to describe 
mono-chromatic incoming electrons are not extremely narrow as 
they should be. If they were infinitely narrow, then we would expect both 
RP results to coincide.

The RP does not change if we use different WPs. We can demonstrate this by using an energetically
narrow wave packet with $\Delta \varepsilon_{WP_{i}}\approx 0.0130$ a.u. 
to compute the RPs and comparing the result with the RPs computed using a
wide WP with $\Delta \varepsilon_{WP_{i}}\approx 0.0255$ a.u. The lower panels in Fig. 11 show that the respective
RP curves compare very well in the energy regions where both curves are valid.

\begin{figure}
\includegraphics[width=0.45\textwidth]{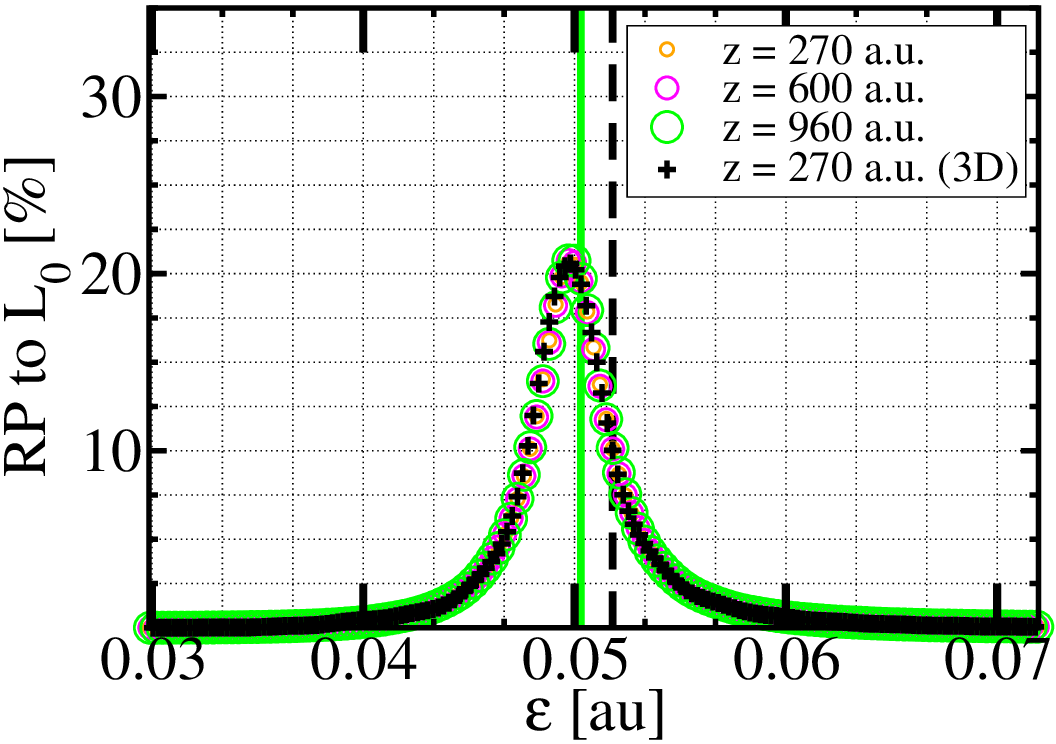}
\caption{\label{surface_plots_C1d}(Color online) Comparison of the ICEC reaction
probability (RP) for configuration (c) of setup C (see Fig. ~\ref{surface_plots_C}) obtained
using the one-dimensional effective model of Sec.~\ref{1dmodel} and the result of 
the full 3D computation. The RP of the full calculation (black crosses) compares 
very well to the one-dimensional results (circles) even for very large grids. The vertical black 
dashed line indicates the value of $\varepsilon^{(peak)}_i$ given by Eq.~(\ref{e_peak})
and the vertical full green line the $\ket{L_{0}L_{1}}$ resonance energy.}
\end{figure}

The comparison of the full 3D and the one-dimensional model for 
setup C is shown in Fig.~\ref{surface_plots_C1d}. We chose the parameters 
of configuration (c) of Fig.~\ref{surface_plots_C}, where
the greatest RP due to resonance-enhanced ICEC occurs. 
As for setup B, the evolution of the electronic density is very similar 
to that of  Fig.~\ref{surface_plots_C}(c) and the RP is almost identical.

The one-dimensional RPs were computed for three different grid lengths, and no difference with the
results of the computed full 3D RPs ($270$ a.u.) is observed for grids up to $960$ a.u. This shows that
the RP is a robust and reliable quantity which is independent of the WP used 
and, to a great extent, also of the grid size. This important point 
is further discussed below.

In principle, one could estimate the RP for a given energy by studying the 
populations $P_{L_{0}}$ or $P_{R_{0}}$ of the one-electron states $L_0$ and $R_0$ 
of the left and right QDs computed using an energetically narrow WP and a
long grid. For setup B this estimate works well as we did use a narrow WP. 
For setup C, however, we used in the full 3D calculations an energetically 
wide WP and a rather short grid and one cannot expect the above mentioned 
estimate to produce realistic results. Indeed, our calculations of these populations
and of the norm $N(t)$ of the wave packet show that these quantities decay due to absorption into the
boundaries of the grid before the estimate takes on the correct value. 
This is mainly because the WP used is very wide. This raises the question 
on why is then the RP computed employing Eq. (19) not affected by the grid size
as is demonstrated in Fig. 12. The answer is that this equation keeps collecting
the flux on the boundary as long as the population $P_{R_{0}}$ on the right QD decreases and that of the left
QD, $P_{L_{0}}$, increases (see Eqs.~(\ref{scatt_matrix}-\ref{g}). 
Clearly, absorption on the boundaries does not affect the RP of ICEC when
computed via these equations. In other words, the RP is very robust against
absorption and this also explains the insensitivity of the results to the size
of the grid and width of the wave packet as found above.

The results show that the overall density evolution is very
similar and the 1D model provides very good results for the RP in both setups 
B and C. Moreover, sometimes it is only possible to perform one-dimensional computations using grids long
enough to show the complete ICEC process.
This assertion strongly supports the use of one-dimensional effective
models when $\varepsilon_i$ is low and thus is not able to produce
excitations in the lateral confinement. The one-dimensional model is a very 
useful tool if the RPs of many different configurations needs to be 
analyzed, because it allows to quickly identify the relevant 
energy range and shape of the RPs.

\section[discussion]{Discussion}\label{discussion}

We demonstrated that ICEC is operative and in some cases a very
effective electron capture mechanism in DQDs.  In the previous
sections we have shown how a simple full-dimensional model can be 
constructed 
to describe the process. Nevertheless, our model includes
only electron correlation to mediate electron capture, 
although other capture mechanisms are likely to be as 
effective as ICEC. Therefore we stick to an estimation on the importance of ICEC 
with respect to other
processes. As we will show, the capture times for ICEC are in the same order or 
even faster than other common mechanisms.  

The capture rate into QDs is the commonly used quantity to characterize
the efficiency of an electron capture process and it depends 
strongly on the amount of time
it takes for the capture to be completed, \emph{i. e.} a faster capture leads to
a greater efficiency. The importance of ICEC is then determined by 
comparing the time it takes ICEC
to complete capture compared to the 
electron capture times reported for 
other processes available in the system.~\cite{prasankumar_ultrafast_2009, 
porte_ultrafast_2009, 
sauvage_long_2002, robel_exciton_2006}

To estimate the speed of ICEC we transfer the parameters of our model to 
realistic
semiconductor structures using the effective mass conversion 
of Table~\ref{tab:H_par}. It is applicable to 
gate defined DQDs with quasi-one dimensional 
geometry~\cite{fasth_tunable_2005,fujisawa_spontaneous_1998} 
or to QDs embedded in nanowires,~\cite{salfi_electronic_2010} so we compare ICEC 
times with those 
obtained for other capture processes in these systems. 
Table~\ref{tab:real_par_b}
shows the energies and sizes for different materials in setup B, 
case (a) and
Table~\ref{tab:real_par_c} those for setup C, case (c). The 
energies obtained are 
well in the range of intraband level spacings of QDs
in nanowires~\cite{salfi_electronic_2010,roddaro_manipulation_2011} and of 
intrashell levels in self-assembled QDs.~\cite{zibik_long_2009}

Let us first analyze setup B. The time window shown in 
Fig.~\ref{surface_plots_B} is about $T=1400$ a.u. and by 
transforming to 
SC materials of Table~\ref{tab:real_par_b} we obtain
$T^{GaAs}=77.8$, $T^{InP}=71.3$, $T^{AlN}=6.1$,  $T^{InAs}=267.3$ ps.
 The time it takes the ICEC process to capture and emit the 
electron can be estimated {from the reaction 
probability if we take into account the time-energy 
uncertainity and the fact that the process gives a peak-shaped 
RP. The RP line shape 
can then be fitted to a Breit-Wigner resonance line shape. We 
performed such a fitting and 
find for case (a) $t^{RP}_{ICEC}=28$ au, and the times in 
different 
materials are accordingly:} {$t^{GaAs}_{ICEC}=1.6$, 
$t^{InP}_{ICEC}=13.1$, 
$t^{AlN}_{ICEC}=0.12$ and 
$t^{InAs}_{ICEC}=5.45$ ps. We stress that this time estimation
only makes sense because we obtained a resonant behavior, rather than a non 
zero contribution for all energy values.}

The surprisingly short time scale it takes ICEC to occur makes ICEC a 
promising mechanism competitive with other capture processes. 
It is faster than the reported capture times of $100$ ps for free carriers in 
bulk GaAs into InAs/GaAs QDs in single layer samples measured at 
room temperature~\cite{turchinovich_inas/gaas_2003}.

The time scale of ICEC obtained for the different geometries always gives 
shorter times for smaller sizes of the DQD. This fact stresses the importance of 
confinement for the process to be competitive. It can be connected to 
previous studies on ICD in molecular dimers, where the length scale 
of about $0.3$ nm typically corresponds to lifetimes in the range of 
several fs.~\cite{cederbaum_giant_1997}

\begin{table}
\centering 
\caption{Realistic values of the parameters in different semiconductors 
for geometry (a) in setup B. The energies are given
in meV and the lengths in nm. Effective masses and dielectric constants taken from~[\onlinecite{singh_physics_1993,levinshtein_properties_2001}]}.
\begin{ruledtabular}
\begin{tabular}{  c | c | c | c | c } 
Parameter & GaAs & InP & AlN & InAs \\
\hline
 $R$ & $97.94$ & $89.89$ & $11.24$ & $286.15$  \\
 $\frac{1}{\sqrt{b_{R/L}}}$ & 9.79 & 8.99 & 1.12 & 28.61  \\
 $l$ & 3.86 & 3.70 & 1.58 & 6.08  \\
 $V_{L}$ & 9.48 & 4.87 & 120.52 & 2.76  \\
 $V_{R}$ & 7.11 & 7.75 & 90.39 & 2.07  \\
 $E_{L_0}$ & -4.47 & -4.87 & -56.78 & -1.30  \\
 $E_{R_0}$ & -2.92 & -3.18 & -37.10 & -0.85  \\
\end{tabular}
\end{ruledtabular}
\label{tab:real_par_b} 
\end{table}

\begin{table}
\centering 
\caption{Realistic values of the parameters in different semiconductors for 
geometry (c) in setup C. The energies are given
in meV and the lengths in nm. Effective masses and dielectric constants taken from~[\onlinecite{singh_physics_1993,levinshtein_properties_2001}].}
\begin{ruledtabular}
\begin{tabular}{  c | c | c | c | c } 
Parameter & GaAs & InP & AlN & InAs \\
\hline
 $R$ & $97.94$ & $89.89$ & $11.24$ & $286.15$  \\
 $\frac{1}{\sqrt{b_{L}}}$ & 17.88 & 16.41 & 2.05 & 52.24  \\
 $\frac{1}{\sqrt{b_{R}}}$ & 9.79 & 8.99 & 1.12 & 28.61  \\
 $l$ & 3.86 & 3.70 & 1.58 & 6.08  \\
 $V_{L}$ & 8.42 & 9.17 & 106.96 & 2.45  \\
 $V_{R}$ & 7.11 & 7.75 & 90.39 & 2.07  \\
 $E_{L_0}$ & -5.23 & -5.69 & -66.40 & -1.52  \\
 $E_{R_0}$ & -2.92 & -3.18 & -37.10 & -0.85  \\
 $E_{L_1}-E_{R_0}$ & 0.57 & 0.62 & 7.19 & 0.16  \\
\end{tabular}
\end{ruledtabular}
\label{tab:real_par_c}
\end{table}

For the setup C case (c) the time window shown in 
Fig.~\ref{surface_plots_C} is of {$T=2700$} a.u. and transforming it 
to the semiconductor 
materials of Table~\ref{tab:real_par_b} we obtain
$T^{GaAs}=150.0$ , $T^{InP}=137.5$ , $T^{AlN}=11.8$ ,  $T^{InAs}=515.6$ ps. 
We can in this setup estimate the duration of the emission using the lifetime 
of the involved resonance $\ket{L_1R_0}$. We have that for case (c) $\tau = 256.8$ a.u. 
which gives the following times in real semiconductors $\tau^{GaAs}=14.26$, $\tau^{InP}=13.08$, 
$\tau^{AlN}=1.12$, $\tau^{InAs}=49.04$ ps. 
From the observed values of the GaAs energy spacings and electron energies in 
the range of $< 5$~meV, 
the decay of the L$_1$R$_0$ resonance in ICEC seems to be competitive
with relaxation via phonons. The times for ICEC are, however, faster than
reported intraband decay times due to acoustic phonon emission for InGaAs/GaAs
QDs of $100$ ps.~\cite{zibik_long_2009}

Our work is focused on strongly laterally 
confined structures, such as nanowires, and is thus suitable for the use of a 
one-dimensional effective potential. In all cases and setups treated here both 
the full and one-dimensional descriptions provided almost identical qualitative and quantitative results. 
The main result obtained from this comparison for the cases studied in 
this work is that the physics in the strongly laterally confined model 
can be correctly described using the effective potential when the 
characteristic lateral energies are about twice or more than those of the QDs.

\section[conclusion]{Conclusions} \label{conclusion}

Ultrafast electron capture in single QDs is 
an extensively studied topic nowadays~\cite{nozik_semiconductor_2010,
prasankumar_ultrafast_2009, porte_ultrafast_2009} due to its relevance in the
development of a wide variety of technological applications.~\cite{prasankumar_ultrafast_2009, porte_ultrafast_2009,
narvaez_carrier_2006} As shown here, electron capture 
via the ICEC processes, in which the neighboring QD in a DQD is 
getting ionized, is particularly fast and
can play a significant role in the dynamics contributing to the energy transfer
between QDs. The ICEC mechanisms in DQDs could, in principle, be exploited to be
implemented in devices which generate a nearly monochromatic low energy electron in a
given direction. 

The implementation of DQDs in nanowires using
materials with long carrier lifetimes such as
InP~\cite{prasankumar_ultrafast_2009,roddaro_manipulation_2011} should be favorable for ICEC.
The rate at which the electron capture occurs varies with material and radius of the wire. 
Reported times for carrier trapping 
cover a large range from fast values of $10$ ps for GaAs~\cite{parkinson_transient_2007} 
and $160$ ps for ZnO~\cite{prasankumar_ultrafast_2009} to very 
slow ones such as $1$ ns for InP nanowires.~\cite{titova_dynamics_2007} Using
wires with long carrier trapping times are favorable for ICEC to be active.

The process is driven by long-range Coulomb interactions, so we expect ICEC to 
be also applicable to other QDs geometries like, \emph{e.g.}, self-assembled vertically stacked
dots.~\cite{muller_electrical_2012, benyoucef_single-photon_2012,
porte_ultrafast_2009, zibik_long_2009}

We have derived an effective one-dimensional approach that correctly describes the dynamics and RPs
of all the cases we have considered. This approach reduces considerably the computational efforts and
also demonstrates, by comparison with full 3D computations, that the physics involved is described
correctly by a one-dimensional model as long as the characteristic confinement energy is about twice or
more than that of the QD.

The calculations presented were performed for the same distance $R$ 
between the 
dots. Since long-range correlation is involved in ICEC a rather pertinent
question is how the reaction probability changes with $R$. The
answer has been partially given in the first publications on ICEC in atoms and 
molecules (see 
Ref.~\onlinecite{gokhberg_interatomic_2010}) and for the 
related ICD decay (see Refs.~\onlinecite{bande_dynamics_2011} 
and~\onlinecite{bande_electron_2013}). 
The ICEC cross section has an asymptotic $1/R^6$ decay with the distance, 
according 
to previous theoretical estimates for atoms and molecules. However, there are 
important contributions not considered in the asymptotic formulas leading to 
$1/R^6$ which are due to orbital overlap (see, 
Ref.~\onlinecite{bande_dynamics_2011} for ICD in QDs 
and Ref.~\onlinecite{averbukh_mechanism_2004} for molecules). These 
contributions 
can lead in some cases to 
a much faster ICD process. Furthermore, the quasi-one dimensional geometry of 
the dots considered here has a clear influence on ICD 
(Ref.~\onlinecite{bande_dynamics_2011}) and probably 
also on ICEC. The calculations are rather 
cumbersome and at the moment there is no exhaustive analysis of this kind for 
ICEC, but it will be done in the future. 

\section[acknowledgments]{Acknowledgments}
F. M. P. acknowledges financial support by Deutscher Akademischer
Austauschdienst (DAAD) and Consejo Nacional de Investigaciones
Cient\'{i}ficas y T\'{e}cnicas (CONICET) and A. B. by Heidelberg University
(Olympia-Morata fellowship) as well as Volkswagen foundation (Freigeist
fellowship). L.S.C. and A.B. thank the Deutsche Forschungsgemeinschaft (DFG)
for financial support.


%

\end{document}